\documentstyle[twocolumn,prl,epsfig,aps]{revtex}
\begin{document}
\twocolumn[
\hsize\textwidth\columnwidth\hsize\csname@twocolumnfalse\endcsname
\draft \preprint{To be published in Phys. Rev. Lett.}
\title{Inelastic Light Scattering by Gap Excitations of Fractional
Quantum Hall States at $1/3 \leq \nu \leq 2/3$.}

\author{Moonsoo Kang,$^{1,2}$ A. Pinczuk,$^{1,2}$ B.S. Dennis,$^2$
M.A. Eriksson,$^2$ L.N. Pfeiffer,$^2$ and K.W. West$^2$}
\address{$^1$Departments of Physics and of Applied Physics,
Columbia University, New York, NY  10027}
\address{$^2$Bell Labs, Lucent Technologies, Murray Hill, NJ  07974}
\date{\today}
\maketitle

\begin{abstract}
We report observations of collective gap excitations of the
fractional quantum Hall (FQH) states at filling factors
$\nu=p/(2p+1)$ ($p$=integer), for $1/3 \leq \nu \leq 2/3$, by
inelastic light scattering. The collective gap energies at $\nu=$
1/3, 2/5 and 3/7 show a drastic decrease as the value $\nu =1/2$
is approached. These energies and the one at $\nu=3/5$ display the
linear scaling with $(e^2/\epsilon l_{o})/|2p+1|$ that is
characteristic of composite fermions in Chern-Simons gauge fields.
In a narrow range of $\nu$ centered at 1/2, where the FQH gaps
collapse, we observe a new excitation mode which exists only at
temperatures below 150~mK.
\end{abstract}

 ]

The ground states of 2D electron systems in the regime of the
fractional quantum Hall effect (FQHE) are described as
incompressible quantum liquids with behaviors dictated by
fundamental interactions \cite{large,FQHE}. Strong electron
correlation in the lowest Landau level in the FQHE at filling
factors $\nu=p/(2p+1)$ ($p$=integer) is often described in terms
of composite fermions (CF). These are weakly interacting
quasiparticles in which two flux quanta are attached to each
electron \cite{jain}. CF quasiparticles move in effective
perpendicular magnetic fields $B_{eff}$ = B - $B_{1/2}$, where B
is the perpendicular component of applied field and $B_{1/2}$ is
the field at $\nu$ = 1/2. At $\nu = 1/2$ composite fermions
experience vanishing $B_{eff}$ and the low energy dynamics of CF
quasiparticles resembles that of a liquid of electrons in zero
magnetic field \cite{hlr,ksc}.
\par
The incompressible FQH liquids have collective gap modes that are
charge-density excitations associated with neutral
quasiparticle-quasihole pairs \cite{bert,haldane85,girvin85}. The
modes have wave vector dispersions determined by interactions
between the quasiparticles. Characteristic features due to
interactions are the rotons, or magnetorotons, at wave vectors $q
\approx$ $1/l_{o}$, where $l_{o}$ = $\sqrt{\hbar c/eB}$ is the
magnetic length.  The energies of non-interacting
quasiparticle-quasihole pairs are the $q \rightarrow \infty$ gap
excitation energies, which are determined by thermally activated
resistivity \cite{du,manoharan}. The scaling of activation gap
energies with filling factor is consistent with a picture in which
the FQHE arises from the low energy dynamics of CF quasiparticles
moving in a $B_{eff}$.
\par
The composite fermion framework has been employed in extensive
theoretical investigations of the FQH liquid. Analytical studies
and numerical evaluations have explored the frequency and wave
vector dependence of response functions and collective modes of
the lowest Landau level FQHE \cite{frad,simon,xie,kam96}. These
results offer detailed predictions for the energies and wave
vector dispersions of collective gap excitations. Experimental
studies of dispersive collective excitations could test
predictions of CF and Chern-Simons formulations and uncover novel
physics of the FQH liquid.
\par
Inelastic light scattering (ILS) methods offer access to
collective excitations of electrons in the FQH regime. Light
scattering studies of gap modes of the incompressible liquid have
been reported at $\nu = 1/3$ \cite{aron93,aron95,Davies}. The
initial work determined the gap energy in the long wavelength ($q
\approx 0$) limit \cite{aron93}. In subsequent studies a mode at
lower energy was assigned to the critical point at the
magnetoroton minimum in the mode dispersion \cite{aron95,Davies}.
ILS by rotons with relatively large wave vectors ($q \approx
10^{6}$~cm$^{-1}$) was explained by loss of translational symmetry
due to residual disorder. Evidence of magnetoroton gap excitations
is also found in absorption of ballistic acoustic phonons
\cite{mellor}. ILS by collective gap excitations of FQH states
other than $\nu =1/3$ remain largely unexplored.
\par
In this Letter we report the first inelastic light scattering
study of collective gap excitations of several FQH states at
filling factors $\nu=p/(2p+1)$, ($p$=integer), within $1/3 \leq
\nu \leq 2/3$. Non-zero scattering wave vectors $q l_{o} \approx
0.1$ enable the acquisition of light scattering spectra at
relatively low electron densities, $n \approx 5\times
10^{10}$~cm$^{-2}$, even in the presence of intense luminescence.
We note that the dynamical structure factor, the function that
enters in conventional expressions for the scattering
cross-sections, is $S(q,\omega)\sim n(ql_{o})^{4}$ for
$q\rightarrow 0$, a rule that works in favor of lower density
systems that have larger values of $l_{o}\sim 1/\sqrt{n}$.
\par
Collective gap modes of incompressible states with marked
temperature and magnetic field dependence are observed at
fractional fillings $\nu=$1/3, 2/5, 3/7, 2/3, and 3/5. The gap
energies at $\nu=$ 1/3, 2/5, and 3/7 decrease drastically as the
filling factor $\nu = 1/2$ is approached, and suggest a collapse
of the collective excitation gap before $\nu$ reaches 1/2. In this
relatively small range of $\nu$, where the gap of the liquid has
collapsed, we uncover a new collective mode that has a marked
temperature dependence for $T \leq 150$~mK. The mode energy has a
dependence on total magnetic field that suggests a link to the
spin degree of freedom of CF quasiparticles.
\par

We studied the high quality 2D electron system in single GaAs
quantum wells (SQW) of widths $d=330$~\AA\/. We present results
obtained in a sample that has $n = 5.4\times 10^{10}$~cm$^{-2}$.
The low temperature mobility of $\mu=7.2\times 10^{6}~cm^{2}$/Vs
is remarkably high considering its low density.
Samples were mounted on the cold finger of a
${}^{3}\text{He}/{}^{4}\text{He}$ dilution refrigerator that is
inserted in the cold bore of a superconducting magnet with windows
for optical access. Cold finger temperatures were as low as 45~mK.
Light scattering spectra were excited with the emission of an
external cavity tunable semiconductor diode laser. The power
density was kept below $10^{-4}$~W/cm$^2$ to prevent heating of
the electron gas. Incident photon energies $\omega_{L}$ were tuned
close to the fundamental optical gap of the GaAs SQW to resonantly
enhance the light scattering intensities. The back scattering
geometry shown in Fig. 1(a) was used. For $\theta = 30^{\circ}$
and a laser wavelength of $\lambda_{L}\approx$ 815~nm the light
scattering vector is $q=(4\pi /\lambda_{L}) \sin\theta \approx
8\times 10^4$~cm$^{-1}$, which gives $q l_{o} \lesssim 0.1$
\par

\begin{figure}
\begin{center}
\epsfig{file=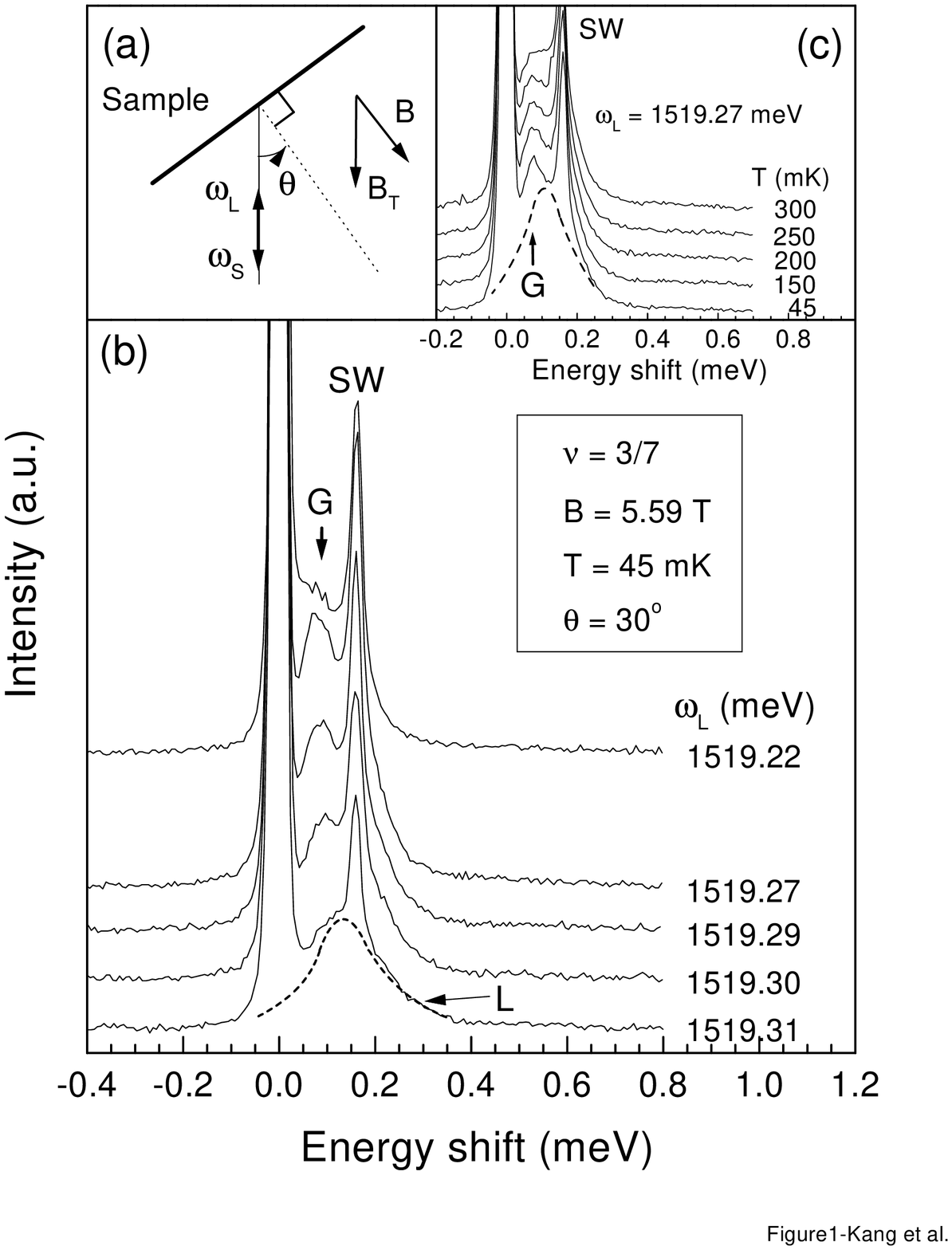,width=2.6in,clip=}
\end{center}
\caption{ (a) Schematic representation of the experimental
geometry. (b) Resonant inelastic light scattering spectra at $\nu
= 3/7$. SW and G denote the long wavelength spin wave and the
collective gap excitations, respectively. The spin wave excitation
is at the Zeeman energy $E_{Z} = g \mu_{B} B_{T}$, where $g = 0.43
\pm 0.01$. The dashed line, which is an approximate guide to the
eye, indicates the luminescence background. (c) Temperature
dependence of the light scattering spectra at $\nu=3/7$. }
\label{fig1}
\end{figure}

Figure~\ref{fig1}(b) shows resonant ILS spectra at $\nu =$ 3/7 and
temperature T=45~mK. $\omega _{L}$ is varied to tune the resonance
enhancement and to distinguish light scattering peaks from
luminescence bands \cite{aron93}. Sharp peaks (FWHM $\leq
0.06$~meV) labeled G and SW are due to ILS by excitations of the
2D electron system. Peak G, at 0.08~meV occurs only in a small
interval $\Delta B \leq 0.1$~T centered at the magnetic field of
$\nu$ = 3/7, and has the marked temperature dependence shown in
Fig.~\ref{fig1}(c). Such pronounced temperature and magnetic field
dependences associate this peak with a collective excitation of
the FQH state. The sharp peak labeled SW is observed over a very
wide range of magnetic field. Its energy is proportional to the
total magnetic field $B_{T}$ and is close to the Zeeman energy
$E_{Z}$ of electrons in GaAs. For this reason we assign it to the
long wavelength ($q \approx 0$) spin wave (SW) excitation
\cite{bert}. The broader bands, FWHM $\approx$ 0.2~meV, near or
under light scattering peaks, are luminescence due to optical
transitions of the GaAs SQW. The luminescence spectrum in this
range of photon energies is shown as a dashed line, and in ILS
plots the luminescence band shifts as $\omega_{L}$ is changed.
\par

\begin{figure}
\begin{center}
\epsfig{file=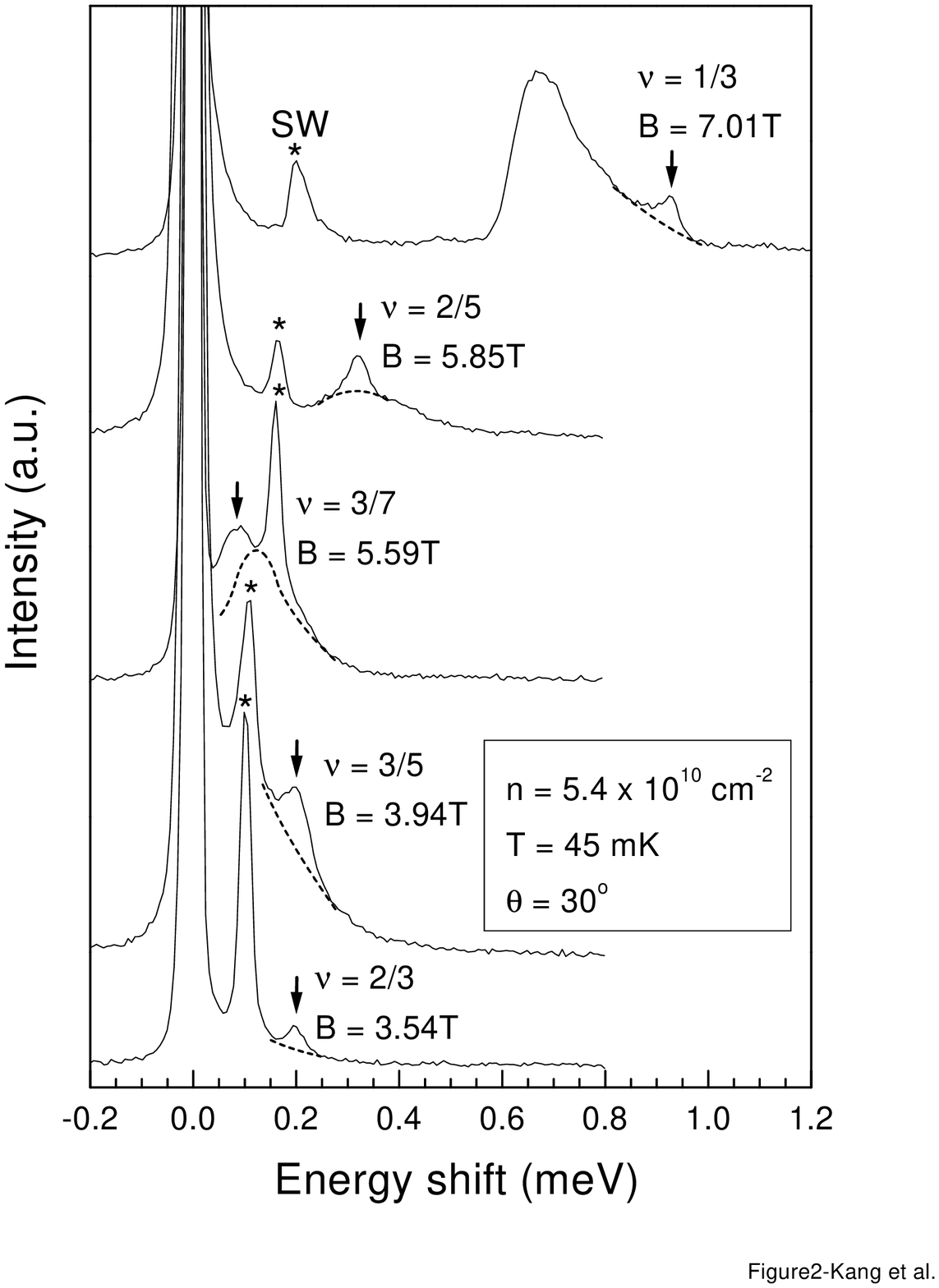,width=2.2in,clip=}
\end{center}
\caption{ Collective gap excitations in the FQH states with
various fractional filling factors within $1/3 \leq\nu\leq 2/3$.
Arrows ($\downarrow$) indicate the gap excitation and SW (*) the
long wavelength spin wave excitation. The spectra shown are
arbitrarily scaled, and the intensity of a given peak depends
strongly on resonance conditions as shown in Fig.~\ref{fig1}(b). }
\label{fig2}
\end{figure}

Figure~\ref{fig2} shows ILS spectra of low-energy collective
excitations measured at several FQH states. In addition to a
spectrum at $\nu$ = 3/7, we show spectra at $\nu$ = 1/3, 2/5, 3/5
and 2/3. While light scattering intensities of the mode measured
at $\nu = 1/3$ persist to temperatures close to 1K, intensities
measured at $\nu$ = 2/5, 3/7/, 3/5 and 2/3 have more dramatic
temperature dependences qualitatively similar to that shown in
Fig.~\ref{fig1}(c). Such dramatic temperature dependences are
typical of FQHE states with $|p| >1$ \cite{du,manoharan}. The
narrow widths (FWHM $\leq$ 0.06~meV) of the excitation modes
indicate that wave vector is conserved in these spectra, and that
we observe long wavelength modes ($q l_{o} < 0.1$). ILS spectra
measured with breakdown of wave vector conservation, such as those
due to magnetorotons, typically have broader spectral shapes (FWHM
$\geq$ 0.15~meV)\cite{aron95}.
\par

\begin{figure}
\begin{center}
\epsfig{file=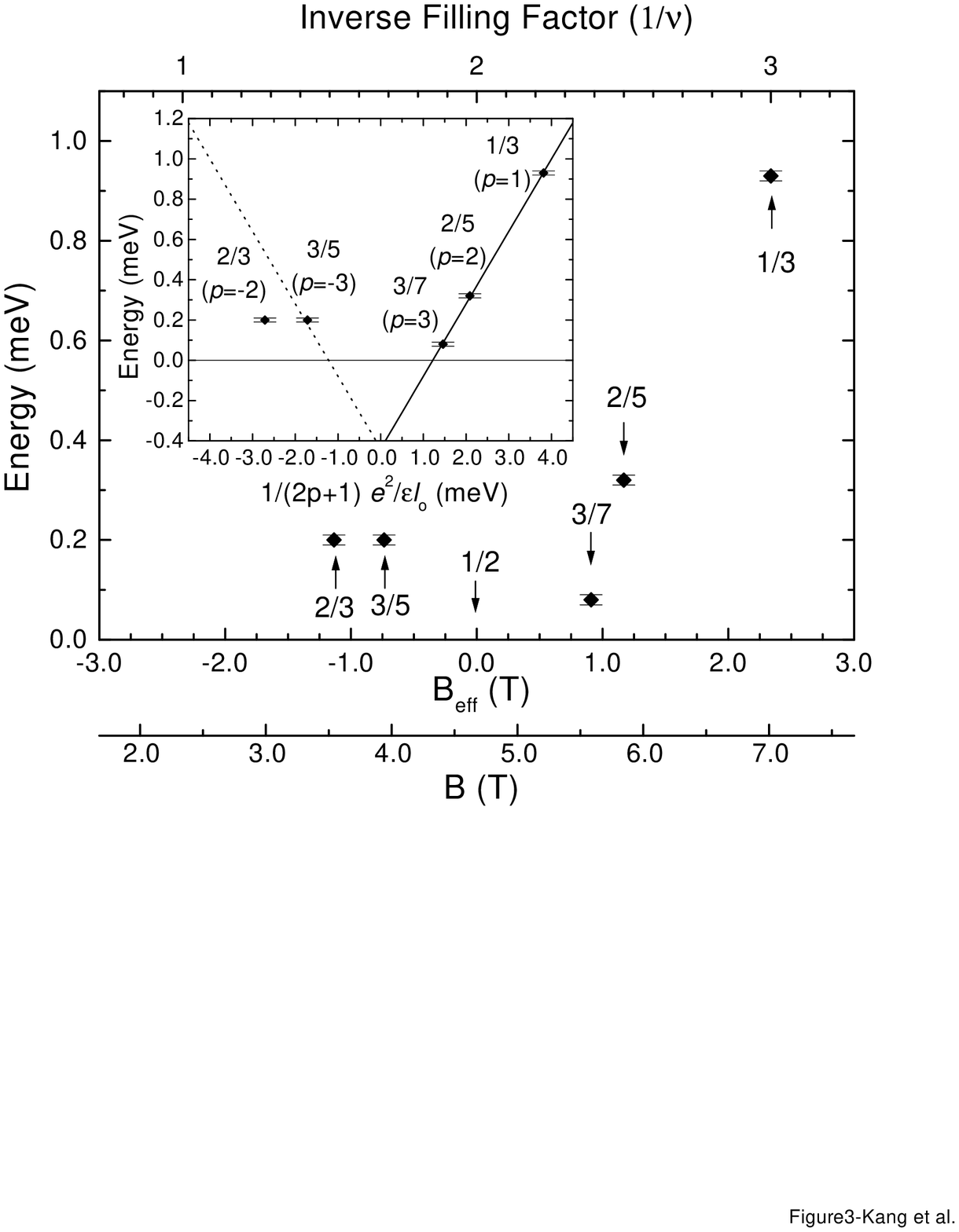,width=3.2in,clip=}
\end{center}
\caption{ The energies of long wavelength FQH state gap. Inset:
The gap energy vs. $(e^2/\epsilon l_{o})/(2p+1)$. The solid line
indicates a linear fit of gap energies at $\nu=$ 1/3, 2/5, and 3/7
to Eq.~\ref{gapeq2} and the dotted line a symmetric image of the
linear fit around $\nu=1/2$. } \label{fig3}
\end{figure}

The modes shown in Figs.~\ref{fig1} and \ref{fig2} are seen only
in narrow ranges of magnetic field centered at the field of the
respective FQH states. The pronounced dependences on $B$ and $T$
identify the modes as gap excitations of the incompressible
states. The energies of these long wavelength gap modes are
plotted as a function of magnetic field in Fig.~\ref{fig3}. The
energies at $\nu=$ 1/3, 2/5, and 3/7 decrease drastically as the
magnetic field B approaches $\nu=$ 1/2. These results suggest that
incompressible FQH states with filling factors very close to 1/2
may be unstable. Trends towards instabilities of the FQHE states
before reaching the compressible state at $\nu = 1/2$ are also
found in measurements of activation gaps\cite{du,manoharan}. At
$\nu=$ 1/3, the measured $q \approx 0$ gap energy in the unit of
Coulomb energy ($E_{c}$ = $e^{2}$/$\epsilon l_{o}$, $\epsilon$ is
the dielectric constant) is about twice the activation gap
energies from Ref.~\cite{du}. At higher fractions, the ratio of
the $q \approx 0$ gap to the activation gap from Ref.~\cite{du}
gets smaller suggesting that $q \approx 0$ gaps measured from our
sample vanish more rapidly as $\nu$ approaches 1/2. However, it
should be noted that the electron density in our sample is about
two to four times lower than in Ref.~\cite{du}.
\par

To interpret these results we recall that the dependence of
activation gaps on $\nu$ has been considered within the CF
framework \cite{hlr,j&k}. The gap energy at $\nu=p/(2p+1)$ is
written as $\Delta(\nu) \sim  \frac{1}{|2p+1|} E_{c}$. The scaling
with $\frac{1}{|2p+1|} E_{c}$ is characteristic of CF
quasiparticles moving in effective magnetic fields that
incorporate Chern-Simons gauge fields \cite {hlr}. This relation,
however, predicts a collapse of the FQH gap exactly at $\nu=1/2$.
The collapse of the FQH gap before the system reaches $\nu=1/2$
was attributed either to the broadening of the fermionic states
due to residual disorder \cite{du,manoharan}, or to the finite
thickness of the 2D electron system \cite{p&j}. The measured
activation gaps are described by the empirical equation
\begin{equation}
\Delta(\nu)=\frac{C}{|2p+1|}\frac{e^2}{\epsilon l_o}-\Gamma,
\label{gapeq2}
\end{equation}
where $\Gamma$ represents the effect of disorder or finite width
of 2D electron systems.
\par

The inset to Fig.~\ref{fig3} shows the results of a fit of the
measured long wavelength gap modes  with Eq.~\ref{gapeq2}, which
reveals that the long wavelength gap energies of FQH states at
$\nu=$ 1/3, 2/5, 3/7 and 3/5 show an excellent scaling with
$\frac{1}{|2p+1|} E_{c}$. It is intriguing that the measured gap
energy at $\nu=2/3$ not only shows a large deviation from the
dotted line but has a value close to the one at $\nu =$ 3/5.
\par

The observation that the long wavelength gap excitation energies
at $\nu =$ 1/3, 2/5, 3/7 and 3/5 are described by Eq.~\ref{gapeq2}
is significant because scaling with $\frac{1}{|2p+1|} E_{c}$ was
introduced to interpret activation gap energies (the collective
gap modes in the $q\rightarrow\infty$ limit) \cite{hlr,p&j,j&k}.
At $\nu=1/3$ the $q\approx 0$ gap excitation has been described as
a two-roton \cite{girvin85,aron95}. However, the character of the
long wavelength gap mode at other FQH states remains largely
unknown. The results in Fig.~\ref{fig3} imply that the gap
energies in the $q\approx 0$ and $q\rightarrow\infty$ limits may
be linked, as proposed in Refs.~\cite{hlr,simon}.
\par

\begin{figure}
\begin{center}
\epsfig{file=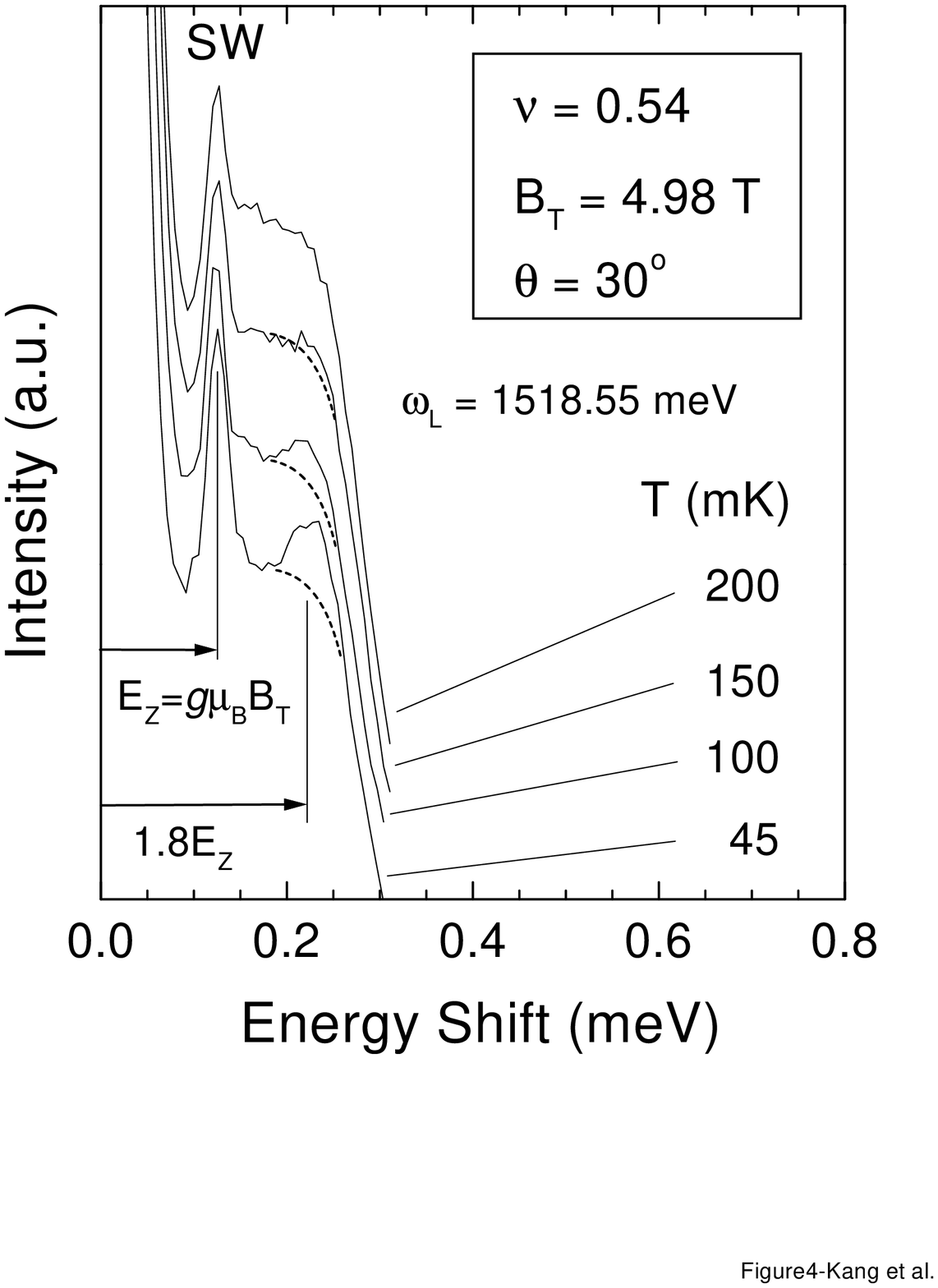,width=2.2in,clip=}
\end{center}
\caption{ Temperature dependence of light scattering at $\nu =
0.54$. Dashed lines denote luminescence backgrounds. }
\label{fig4}
\end{figure}

The results in Fig.~\ref{fig3} also highlight the collapse of the
FQH gaps at filling factors in the range 4/9 $\lesssim \nu
\lesssim$ 4/7, similar to the one first observed in
magnetotransport. We believe it occurs here at smaller values of
$|p|$ because of the lower electron density in our sample. This
implies that in a narrow range of filling factors centered at $\nu
= 1/2$ the long wavelength gap mode is unstable and a FQH liquid
should not exist. ILS spectra obtained in this range of $\nu$
reveal a new collective excitation mode with a remarkable
temperature dependence, as shown in Fig.~\ref{fig4} for $\nu$ =
0.54.
\par

The spectra in Fig.~\ref{fig4} display a sharp ILS peak (FWHM $
\approx 0.06$~meV) that has a temperature dependence for T $\leq$
100~mK and disappears at temperatures T $\geq $ 150~mK. Such
temperature dependence is extraordinary, particularly when
detected in ILS measurements. This excitation is observed at all
investigated magnetic fields within the range in which the FQH
liquid is expected to be unstable. In fact, the range of $\nu$ to
observe the new excitation mode ($0.43\leq\nu\leq 0.57$) coincides
with the range where the gap collapses in the linear fit in the
inset of Fig.~\ref{fig3} ($0.44\leq\nu\leq 0.56$). In this range
of filling factors, the mode energy displays a dependence on total
magnetic field $B_{T}$ that may be represented as 1.8$E_{Z}$. We
note that an energy which is proportional to total magnetic field
is characteristic of an excitation mode associated with the spin
degree of freedom.
\par

Given that the excitation mode appears close to $\nu = 1/2$, when
the FQH states give way to a liquid of CF quasiparticles, we may
conjecture that the mode is characteristic of a novel ground
state. This state emerges at very low temperatures $T \leq
100$~mK, and could be caused by interactions in the CF liquid. We
may conceive a scenario in which a mode at energy 1.8$E_{Z}$ is
constructed as a collective excitation which involves two spin
waves. While a single $q \approx 0$ spin wave excitation is
required to have energy $E_{Z}$ by Larmor's theorem \cite{bert},
the energy shift of the second-order spin excitation from $2
E_{Z}$ could be a manifestation of interactions among CF
quasiparticles. We note that there is extensive theoretical
literature that considers novel correlation effects at $\nu = 1/2$
as well as experimental reports of anomalies at $\nu=$
1/2\cite{willett90,tieke}. Our results suggest that the inelastic
light scattering method could offer an experimental venue to
explore liquid states of CF quasiparticles that emerge near $\nu =
1/2$.

In summary, we measured collective gap excitations of FQH states
at filling factors $1/3 \leq \nu \leq 2/3$ by inelastic light
scattering. The narrow linewidths suggest wave vector conservation
at values $q l_{o} \approx 0.1$. The results indicate an
instability of the FQH states in the vicinity of $\nu=1/2$,
similar to those observed in transport measurements. In a range of
$\nu$ near 1/2, we found a new excitation mode that exists only at
temperatures below 150~mK. Further low temperature studies could
reveal fundamental interactions within the liquid of CF
quasiparticles at filling factors close to $\nu = 1/2$.
\par
We are grateful to B.I. Halperin, J.K. Jain, S.H. Simon, and H.L.
Stormer for many discussions. We are also grateful to K.W. Baldwin
for magnetotransport measurements.

%
%

\begin{references}
\bibitem{large}
R.B. Laughlin, Phys. Rev. Lett. {\bf 50}, 1395 (1983).

\bibitem{FQHE}
For recent reviews see {\em Perspectives in Quantum Hall Effects},
edited by S. Das Sarma and A. Pinczuk (John Wiley \& Sons, New
York, 1997); and
H.L. Stormer, D.C. Tsui, and A.C. Gossard, Rev. Mod. Phys.
{\bf 71}, S298 (1999).

\bibitem{jain}
J.K. Jain, Phys. Rev. Lett. {\bf 63}, 199 (1989); Phys. Rev. B
{\bf 40}, 8079 (1989); {\bf 41}, 7653 (1990).

\bibitem{hlr}
B.I. Halperin {\em et al.}, Phys. Rev. B {\bf47}, 7312 (1993).

\bibitem{ksc}
V. Kalmeyer and S.-C. Zhang, {\em ibid.} {\bf 46}, 9889 1992).

\bibitem{bert}
C. Kallin and B.I. Halperin, {\em ibid.} {\bf 30}, 5655 (1984).

\bibitem{haldane85}
F.D.M. Haldane and E.H. Rezayi, Phys. Rev. Lett. {\bf 54}, 237 (1985).

\bibitem{girvin85}
S.M. Girvin {\em et al.}, Phys. Rev. Lett. {\bf 54}, 581 (1985);
Phys. Rev. B {\bf 33}, 2481 (1986).

\bibitem{du}
R.R. Du {\em et al.}, Phys. Rev. Lett. {\bf 70}, 2944 (1994).

\bibitem{manoharan}
H.C. Manoharan {\em et al.}, Phys. Rev. Lett. {\bf 73}, 3270
(1994).

\bibitem{p&j}
K. Park and J.K. Jain, Phys. Rev. Lett. {\bf 81}, 4200 (1998).

\bibitem{frad}
A. Lopez and E. Fradkin, Phys. Rev. B {\bf 47}, 7080 (1993).

\bibitem{simon}
S.H. Simon and B.I. Halperin, {\em ibid.} {\bf 48}, 17368 (1993);
{\bf 50}, 1807 (1994); S. He, S.H. Simon, and B.I. Halperin, {\em
ibid.} {\bf 50}, 1823 (1994).

\bibitem{xie}
X.C. Xie, {\em ibid.} {\bf 49}, 16833 (1994).

\bibitem{kam96}
R.K. Kamilla {\em et al.}, Phys. Rev. Lett. {\bf 76}, 1332 (1996);
R.K. Kamilla and J.K. Jain, Phys. Rev. B {\bf 55}, 13417 (1997);
V.W. Scarola {\em et al.}, cond-mat/9910491.

\bibitem{aron93}
A. Pinczuk {\em et al.}, Phys. Rev. Lett. {\bf 70}, 3983 (1993).

\bibitem{aron95}
A. Pinczuk {\em et al.}, Bull. Am. Phys. Soc. {\bf 40}, 515
(1995); A. Pinczuk {\em et al.}, Physica B {\bf 249-251}, 40
(1998).

\bibitem{Davies}
H.D.M. Davies {\em et al.}, Phys. Rev. Lett. {\bf 78}, 4095
(1997).

\bibitem{mellor}
C.J. Mellor {\em et al.}, Phys. Rev. Lett. {\bf 74}, 2339 (1995);
U. Zeitler {\em et al.}, {\em ibid.} {\bf 82}, 5333 (1999).

\bibitem{j&k}
J.K. Jain and R.K. Kamilla, Phys. Rev. B {\bf 55}, 4895 (1997).

\bibitem{willett90}
R.L. Willett {\em et al.}, Phys. Rev. Lett. {\bf 65}, 112 (1990).

\bibitem{tieke}
B. Tieke {\em et al.}, Phys. Rev. Lett. {\bf 78}, 4621 (1997).

\end{references}
\end{document}